\documentclass[twocolumn,nofootinbib,floatfix,prb]{revtex4}

\makeatletter
\def\@dotsep{4.5}
\makeatother

\usepackage{dcolumn}
\usepackage{graphicx}

\begin{document}

\title{Application of Ewald summations to long-range dispersion forces}
\author{Pieter J. in 't Veld}
\author{Ahmed E. Ismail} 
\email[Email: ]{aismail@sandia.gov}
\author{Gary S. Grest}
\affiliation{Sandia National Laboratories, Albuquerque, New Mexico 87185}

\begin{abstract}
We present results illustrating the effects of using explicit summation terms for the $r^{-6}$ dispersion term on the interfacial properties of a Lennard-Jones fluid and SPC/E water.
For the Lennard-Jones fluid, we find that the use of long-range summations, even with a short ``crossover radius,'' yields results that are consistent with simulations using large cutoff radii.
Simulations of SPC/E water demonstrate that the long-range dispersion forces are of secondary importance to the Coulombic forces.
In both cases, we find that the ratio of box size $L_{\parallel}$ to crossover radius $r_{\rm c}^{\mathbf k}$ plays an important role in determining the magnitude of the long-range dispersion correction, although its effect is secondary when Coulombic interactions are also present.
\end{abstract}

\maketitle

\section{Introduction}

One of the fundamental problems in molecular simulations is how to sum efficiently the pairwise interaction potential $u(r)$ that forms the basis for calculation of nonbonded interactions.
If the potential satisfies $\left| u(r) \right| < Ar^{-m}$ for $m > 3$, then the sum is guaranteed to be absolutely convergent; otherwise, it is at best conditionally convergent.
Even in cases where the sum is absolutely convergent, however, explicit calculation of all the pairwise interactions, an $O(N^{2})$ operation per time step, quickly becomes computationally unfeasible for molecular dynamics simulations.
Circumventing this difficulty usually involves one of two routes. In the first, a cutoff $r_{\rm c}$ is introduced, outside of which interactions are neglected explicitly during the simulation. Long-range contributions to chemical potential, surface tension, and other thermodynamic properties are calculated by integrating from the cutoff radius $r_{\rm c}$ to infinity, assuming the radial distribution function $g(r)$ equal to unity.
Certain cases, however, require large cutoff radii, which leads to computational inefficiency because of the large number of pairs needed to calculate the total pairwise sum.

An alternative method involves transformation of the pairwise interactions beyond some ``crossover'' distance (which we designate as $r_{\rm c}^{\mathbf{k}}$) into a more readily-solved problem.
Although truncation of the pairwise potential is more computationally efficient, and has in fact been used in parametrization of the SPC/E, \cite{Berendsen87} TIP3P,\cite{Jorgensen83} and TIP4P\cite{Jorgensen83} models of water, excluding interactions from consideration in the simulation can introduce significant errors in the values of thermodynamic properties obtained from a simulation.
In such cases one generally relies on Ewald summations, a technique which replaces the long-range calculations by a sum of replicas of the central box. 
We consider the effects of including explicit summation of long-range dispersion forces.
These long-range forces are normally neglected in one-phase systems, but become of importance in multiphase systems or when studying interface phenomena.

Following earlier work by Williams \cite{Williams71} and Perram and co-workers,\cite{Perram89} Karasawa and Goddard used long-range summations of dispersion forces to derive thermodynamic properties of argon and of NaCl crystals. \cite{Karasawa89}
More recently, several other investigations into the use of long-range summation of dispersion forces have been reported.
Ou-Yang {\em et al}.\ have used long-range summation to study the phase equilibrium of Lennard-Jones fluids. \cite{Ouyang05}
L\'opez-Lemus and co-workers have used Ewald summation to study the density of $n$-alkanes. \cite{Lopez06} 
Shi {\em et al}.\ have extended the particle-particle particle-mesh (PPPM) implementation of Hockney and Eastwood\cite{Hockney88} to include Ewald sums for the dispersion term, and used it to study the phase envelope of water; \cite{Shi06}
Essmann and co-workers have created a similar mesh-based formulation for Ewald summations.\cite{Essmann95}
Greater emphasis has been placed on how to use long-range corrections, as applied to, for example, constant-pressure \cite{Lague04} and inhomogeneous \cite{Janecek06} simulations.
Deserno and Holm considered the effects of various parameters involved in performing the Ewald summation on its accuracy for several mesh implementations, \cite{Deserno98,Deserno98b} while Pollock and Glosli have evaluated the efficiency of both traditional Ewald summations and mesh implementations. \cite{Pollock96}

In this paper, we present results on the use of Ewald summations to handle the $r^{-6}$ dispersion term.
We first illustrate the effect of incorporating the long-range sum by studying a model Lennard-Jones fluid; we report the liquid-vapor coexistence data, as well as the surface tension.
Next, we demonstrate the effects of long-range summation of the dispersion term in ionic systems by simulating SPC/E water.
In both cases, we compare the Ewald simulation results to results of simulations with truncated Lennard-Jones potentials.
We also examine the role long-range corrections of the surface tension plays in both types of simulations.

In Section \ref{sec2}, we briefly outline the equations that must be implemented to evaluate the Ewald sum for potentials of the form $f(r) \sim Ar^{-6}$.
We then demonstrate its application to a Lennard-Jones fluid and to water, which has both Coulombic and dispersion Ewald sums, in Sections \ref{secM} and \ref{secRD}. 
We consider issues related to the performance of the various algorithms for performing long-range summations in Section \ref{secP}.
Finally, we offer our conclusions in Section \ref{secC}.

\section{Mathematical Formulation of the Ewald Dispersion Sum}
\label{sec2}

Incorporation of long-range summation of LJ potentials,
\begin{equation}
u_{ij}(r) = 4\epsilon_{ij} \left[ \left(\frac{\sigma_{ij}}{r}\right)^{12} - \left(\frac{\sigma_{ij}}{r}\right)^{6} \right] = \frac{A_{ij}}{r^{12}} - \frac{B_{ij}}{r^{6}},
\label{eqnLJ}
\end{equation}
requires treatment of the second, attractive term; the repulsive $r^{-12}$ term can be neglected, particularly as $r$ becomes large.
We now briefly summarize the equations governing long-range summation of the attractive $r^{-6}$ or dispersion portion of the Lennard-Jones potential. 
Note that the following also applies to any other potential with attractive $r^{-6}$ contributions, such as a Buckingham exp-6 potential, provided that excluded volume interactions fall off fast enough.

Karasawa and Goddard \cite{Karasawa89} derive expressions for the dispersion part of energy, force, and stress using Ewald summations.
For the energy they find
\begin{eqnarray}
E_{6} &=& \frac{1}{2\eta^{6}} \sum_{L,i,j} B_{ij}\left(1+a^{2}+\frac{a^{4}}{2}\right) \frac{e^{-a^{2}}}{a^{6}}\nonumber \\
&& {} + \frac{\pi^{2/3}}{24V} \sum_{\mathbf{h \ne 0}} h^{3}\left(\pi^{1/2} \mathrm{erfc}(b) + \left(\frac{1}{2b^{3}} - \frac{1}{b}\right) e^{-b^{2}}\right) \times\nonumber \\
&& {} \times S_6\left(\mathbf{h}\right) S_6\left(\mathbf{-h}\right) \nonumber \\
&& {} + \frac{\pi^{2/3}}{6V\eta^{3}} \sum_{i,j} B_{ij}
- \frac{1}{12\eta^{6}} \sum_{i} B_{ii},
\label{eqnEnergy}
\end{eqnarray}
where $V$ represents the volume of the simulation cell, $\mathbf{h}$ the reciprocal lattice vector, $h = |\mathbf{h}|$, $a = |\mathbf{r}_{i}-\mathbf{r}_{j}-\mathbf{R}_{L}|/\eta$, and $b = h\eta/2$.
The first contribution to the energy sums over lattices $L$ and particles $i$ and $j$.
The parameter $\mathbf{R}_L$ represents the lattice translation vector.
The parameter $\eta$ represents the range of interactions handled in reciprocal space; for a given value of $r$, the larger the value of $\eta$, the more interactions that are included in reciprocal space. 
We define the quantity $r_{c}^{\mathbf k}$ as the ``crossover'' which represents the maximum distance for which interactions are summed using the first term in Eq.\ \ref{eqnEnergy}.
Note that $r_{c}^{\mathbf k}$ in combination with the number of reciprocal lattice vectors  $n_{vec}^{\mathbf k}$ determine the numerical accuracy with which the lattice-sum Hamiltonian is evaluated.
Using a finite $n_{vec}^{\mathbf k}$ in combination with a finite $r_{c}^{\mathbf k}$ results in an approximate solution for the lattice sum.
Conversely, the solution is exact when either one is infinite.
The excluded volume uses the same cutoff as the crossover radius, so no additional cutoff for the $r^{-12}$ term is needed.
The structure factor $S_{6}(\mathbf{h})$, a complex number, is defined as
\begin{equation}
S_{6}(\mathbf{h}) = \sum_{j}b_{j}\exp(-i\mathbf{h \cdot r}_{j}),
\label{eqnStructure}
\end{equation}
assuming geometric mixing of dispersion constants $\left( B_{ij} = \sqrt{B_{ii}B_{jj}} = b_{i}b_{j}\right)$.
This type of mixing applies to OPLS potentials by Jorgensen {\em et al}. \cite{Jorgensen88}
On the other hand, potentials like CHARMM \cite{Mackerrel98} use arithmetic mixing for sigma cross-terms $\left( \sigma_{ij} = \frac{1}{2}\sigma_{ii} + \frac{1}{2} \sigma_{jj} \right)$, in which case it is necessary to expand the dispersion constant,
\begin{equation}
B_{ij} = 4\epsilon_{ij}\sigma_{ij}^{6} = \sum_{k=0}^{6} b_{i,k}b_{j,6-k} ,
\label{eqnExpansion}
\end{equation}
with $b_{i,k} = \frac{1}{4}\sigma_{ii}^{k}\sqrt{c_{k}\epsilon_{ii}}$ and $c_{k} = {6 \choose k}$. This redefines the structure factor product as
\begin{equation}
S_{6}(\mathbf{h})S_{6}(\mathbf{-h}) = 2 \sum_{k=0}^{3} S_{6,k}(\mathbf{h})S_{6,6-k}(\mathbf{-h}), \label{eqnProduct}
\end{equation}
with
\begin{equation}
S_{6,k}(\mathbf{h}) = \sum_{j}b_{j,k}\exp(-i\mathbf{h \cdot r}_{j}) .
\label{eqnSingle}
\end{equation}
Note that symmetry requires the calculation of only four out of the seven structure factors products in Equation \ref{eqnProduct}. The force on atom $k$ is given by
\begin{widetext}
\begin{eqnarray}
\mathbf{f}_{6,k} &=& \frac{1}{\eta^{-8}}\sum_{L,j} B_{kj}(\mathbf{r}_{k}-\mathbf{r}_{j}-\mathbf{R}_{L})
(6a^{-8} + 6a^{-6} + 3a^{-4} + a^{-2})\nonumber \\ 
& & {} + \frac{\pi^{2/3}}{12V} {\rm Im}
\left( \sum_{\mathbf{h \ne 0}} ib_{k}\exp(-i\mathbf{h \cdot r}_{k})h^{3}
\left[\pi^{1/2} {\rm erfc}(b) + \left(\frac{1}{2b^{3}} - \frac{1}{b}\right) e^{-b^{2}}\right] S_{6}(\mathbf{h})\mathbf{h}\right) .
\label{eqnForce}
\end{eqnarray}
\end{widetext}
The instantaneous stress is given by
\begin{widetext}
\begin{eqnarray}
V\Pi_{\alpha \beta} &=& \frac{1}{2\eta^{8}}\sum_{L,i,j} B_{ij}
(6a^{-8} + 6a^{-6} + 3a^{-4} + a^{-2})e^{-a^{2}}
(\mathbf{r}_{i}-\mathbf{r}_{j}-\mathbf{R}_{L})_{\alpha}(\mathbf{r}_{i}-\mathbf{r}_{j}-\mathbf{R}_{L})_{\beta}
\nonumber \\ 
& & {} + 
\frac{\pi^{2/3}}{24V}
\sum_{\mathbf{h \ne 0}} 
\left[ \left( \pi^{1/2}~ {\rm erfc}(b) + \frac{(1-2b^{2})e^{-b^{2}}}{2b^{3}} \right) 
h^{3} \delta_{\alpha\beta} \right] 
S_{6}(\mathbf{h}) S_{6}(-\mathbf{h})
\nonumber
\\
& & {} +
\frac{\pi^{2/3}}{24V}
\sum_{\mathbf{h \ne 0}} 
\left[3h\left( \pi^{1/2}~ {\rm erfc}(b) - \frac{e^{-b^{2}}}{b} \right) 
\mathbf{h_{\alpha}h_{\beta}} \right] 
S_{6}(\mathbf{h}) S_{6}(-\mathbf{h})
\nonumber
\\
&& {} + \frac{\pi^{2/3}}{6\eta^{3}V}\sum_{i,j} B_{ij}\delta_{\alpha\beta}.
\label{eqnStress}
\end{eqnarray}
\end{widetext}

\section{Methodology\label{secM}}

\subsection{Lennard-Jones fluid}

All molecular dynamics simulations were performed using the LAMMPS simulation package \cite{Plimpton95} with added Ewald dispersion summations. 
All Lennard-Jones parameters can be expressed in terms of the distance $\sigma_{ij} = \sigma$, the root of the Lennard-Jones potential (\ref{eqnLJ}); $\epsilon_{ij}$, the well-depth of the potential; and the time, $\tau = \sigma (m/\epsilon)^{1/2}$, where $m$ is the mass of the atom. %
Initial configurations were created by randomly placing either $N_{particles}$ = 4000 or 16000 particles in simulation boxes with respective geometries of $L_{x} =  L_{y} = L_{\parallel} = 11.01\sigma$ or $22.02\sigma$, and $L_{z} = L_{\perp} = 44.04\sigma $.
Avoidance of system size effects due to large cutoffs motivated the choice of using system sizes of either 4000 or 16000 particles.

\begin{figure*}
\begin{center}
\includegraphics[width=6in]{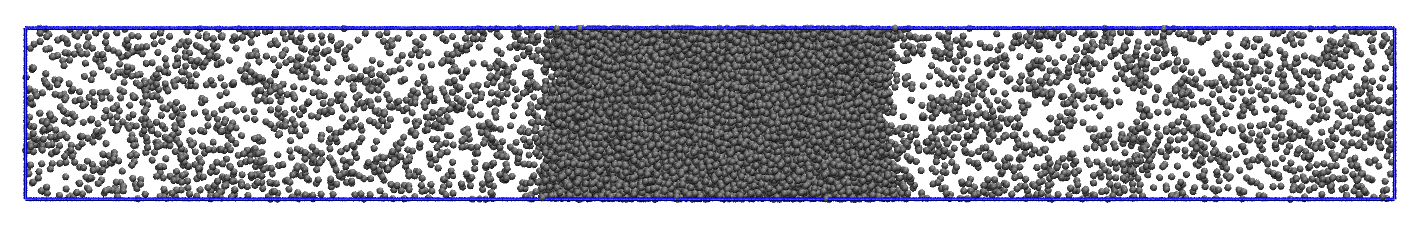}
\caption{Snapshot for a 16000 atom monomeric Lennard-Jones fluid at coexistence with cutoff $r_{c} = 7.5\sigma$ at $T^{*} = 1.10$ shown in the $xz$-plane.\label{Fig0}}
\end{center}
\end{figure*}

Thus created, configurations were equilibrated for 500$\tau$ at $T^* = k_{B}T / \epsilon = 0.7$ with $\Delta t = 0.005 \tau$, using three-dimensional periodic boundary conditions.
Once equilibrated, liquid-vapor interfaces were created by changing the size of the box in the $z$-direction to $L_\perp = 176.16\sigma$ without displacing the particles.
Once strained, the system was equilibrated for $500 \tau$ at $T^* = 0.7$.  Subsequent higher temperatures used this state as initial configuration.
To reach the desired temperature, the system was slowly heated over a period of $500\tau$; the system was then allowed to equilibrate at the final temperature for $250\tau$. 
Figure \ref{Fig0} shows a typical snapshot of a 16000 atom system.

Once at the desired temperature, total system pressure tensors and $z$-directional density profiles were measured every $5\tau$ for 1250 to 5000$\tau$.
All simulations used a Nos\'e-Hoover thermostat with relaxation time 10 $\tau$.
Studied temperatures were $T^* \in \left\{0.70, 0.85, 1.10, 1.20\right\}$.  
Values for $\eta$ follow from the choice of precision through 
$\eta = (1.35 - 0.15 \ln P)/r_c^\mathbf{k}$, where $P$ is the precision.\cite{Pollock96}
Ewald dispersion summations used a precision of $0.05$.  
Changing precisions to 0.01 did not change the results.  
For a given system size, the number of $\mathbf k$-vectors employed in each simulation decreases with increasing $r_{c}^{\mathbf k}$;  the Lennard-Jones simulations described here use between 320 and 1404 $\mathbf k$-vectors.

\subsection{SPC/E water}
The dominant contribution to the surface tension of water is from the electrostatic interactions, which are much stronger than the Lennard-Jones interactions.
However, the traditional Ewald method or the PPPM technique of Hockney and Eastwood \cite{Hockney88} is generally used for the long-range electrostatics only. 
Long-range corrections are applied to the surface tension \cite{Chapela77,Blokhuis95} and other system properties \cite{Allen87} to account for the effects of truncating the Lennard-Jones potential.

The SPC/E water model \cite{Berendsen87} was used for all simulations described below, as we found in a previous study \cite{Ismail06} that the SPC/E model showed acceptable agreement with experimental data while avoiding the computational expense associated with four-point water models. 
The SHAKE technique \cite{Ryckaert77} was used to constrain the bond lengths and bond angles. 
The equations of motion were integrated using the Verlet algorithm; a Nos\'{e}-Hoover thermostat with a relaxation time of 500 fs was used to control the temperature during the simulations.
All simulations used a timestep of $\Delta t = 1.0$ fs.

To determine the effect of the long-range Ewald summation of the dispersion term on the liquid-vapor equilibrium properties of water, water molecules were randomly placed in a periodic, rectangular box. 
In all cases, the dimensions of the box in the directions parallel to the interface were equal: $L_{x} = L_{y} = L_{\parallel}$, where $L_{\parallel}$ varied between $3.65$ nm and $7.74$ nm. 
For all simulations, the dimension perpendicular to the interface was $L_{z} = L_{\perp} = 5.0$ nm. 
The initial density in each box was $\rho = 1.00$ g cm$^{-3}$.

For each different set of Lennard-Jones and electrostatic cutoffs, systems were allowed to equilibrate at 300 K or 400 K for 500 ps. 
The resulting configuration was then placed in a box with $L_{z} = L_{\perp} = 15.0$ nm, and allowed to equilibrate for 1 ns to establish the liquid-vapor interface.
Production runs of at least 1 ns then followed, with pressures measured every timestep and positions stored every 5 ps.

With one exception, for the systems which included long-range summation of the Lennard-Jones dispersion term, the Ewald summation was also used for the electrostatic interactions. 
In such cases, the crossover radius for both the LJ and electrostatic interactions were identical.
For the systems in which the Lennard-Jones potential was truncated, electrostatic interactions were calculated using the PPPM method. \cite{Hockney88} 
The PPPM method yields results that are indistinguishable from Ewald summations (within simulation uncertainty), but have much faster running times for large systems.
The root-mean-squared precision of the algorithm was on the order of $10^{-3}$.
We showed in our previous paper that the interfacial properties of SPC/E water do not show a significant dependence on the accuracy of the $\mathbf{k}$-space mesh, provided that the system is allowed to equilibrate sufficiently.
In our previous studies of the surface tension of water models,\cite{Ismail06} we found that $\mathbf{k}$-space meshes of $5 \times 5 \times 20$ through $20 \times 20 \times 80$ yielded equivalent results. 
The PPPM simulations reported here used meshes of size $10 \times 10 \times 27$; the Ewald summations used a total of 2046 vectors.

\subsection{Surface tension and tail corrections}

There are two primary methods used to compute the surface tension using molecular simulations.
One approach, developed by Tolman \cite{Tolman48} and refined by Kirkwood and Buff,\cite{Kirkwood49} computes the surface tension as an integral of the difference between the normal and tangential
pressures $p_{\perp}\left(  z\right)  $ and $p_{\parallel}\left(  z\right)  $:
\begin{equation}
\gamma_{p}=\frac{1}{2}\int_{-\infty}^{\infty}\left(  p_{\perp}\left(
z\right)  -p_{\parallel}\left(  z\right)  \right)  ~dz, \label{eqn13}
\end{equation}
where, in our geometry,
$p_{\perp}(z)  = p_{z}(z)$, and
$p_{\parallel}(z) = (p_{x}(z)+p_{y}(z))/2$.
Away from an interface, $p_{\perp}=p_{\parallel}$ and the integrand vanishes.
Therefore the nonzero contributions to the integral in Eq.\ (\ref{eqn13}) come from the
interfacial region.
In the case of an interface between two fluid phases, the integral in Eq.\ (\ref{eqn13}) can be replaced with an ensemble average of the difference between the normal and tangential pressures:
\begin{equation}
\gamma_{p}=\frac{L_{z}}{2}\left\langle p_{\perp}-p_{\parallel}\right\rangle
=\frac{L_{z}}{2}\left[  \left\langle p_{z}\right\rangle -\frac{\left\langle
p_{x}\right\rangle +\left\langle p_{y}\right\rangle }{2}\right]  .
\label{eqn14}
\end{equation}
The outer factor of $1/2$ in Eq.\ (\ref{eqn14})\ accounts for the presence of
two liquid-vapor interfaces.

As discussed above, the surface tension measured by simulations, $\gamma_{\rm sim}$, tends to be lower than the experimentally-measured surface tension. 
Improved agreement with experimental data by incorporating the tail correction,
$\gamma_{\rm tail}$:
\[
\gamma = \gamma_{p}+\gamma_{tail},
\label{eqn23}
\]
where $\gamma_{tail}$ can be determined from \cite{Chapela77, Blokhuis95}
\begin{eqnarray}
\gamma_{tail} =
	\frac{\pi}{2} \int_{-\infty}^{\infty} \int_{-1}^{1} \int_{r_{c}}^{\infty}
		r^{3}\left( 1-3s^{2} \right)  \frac{{\rm d}U(r)}{{\rm d}r} g\left(r\right) \times 		\nonumber\\
		\left( \rho \left( z \right) \rho \left( z-sr \right) -
			\left( \rho_{G}\left( z\right) \right) ^{2} \right)
	~{\rm d}r~{\rm d}s~{\rm d} z.
\label{eqn15}
\end{eqnarray}
In Eq.\ \ref{eqn15}, $U\left( r\right) $ is the pairwise potential, $g\left( r\right)$ is the radial distribution function, $\rho\left( z\right)$ is the observed interfacial profile, and $\rho_{G}\left(z\right)$ is a Gibbs dividing surface:
\begin{equation}
\rho_{G}\left( z\right) =
	\rho_{c} + \frac{\Delta \rho}{2} {\rm sgn} \left( z \right) .
\label{eqn16}
\end{equation}
Eq.\ \ref{eqn15} represents the contribution to the stress from the region $r \ge r_{c}$ which is not captured by simulation; the term containing the product of densities is an approximation for the true pair density $\rho^{(2)}$ required by the Kirkwood-Buff theory.\cite{Kirkwood49}
Following our earlier work on tail corrections in the surface tension of water models, we fit the density profile to an error function profile to simplify the calculations.\cite{Ismail06}

\section{Results and Discussion\label{secRD}}

\subsection{Lennard-Jones fluid}

Surface tensions at interfaces in two-phase systems are sensitive to long-range interactions, even when simple Lennard-Jones interactions are concerned.
This phenomenon becomes especially pronounced close to the critical point, where large fluctuations in density dominate.

\begin{figure}
\begin{center}
\includegraphics[width=3.2255in]{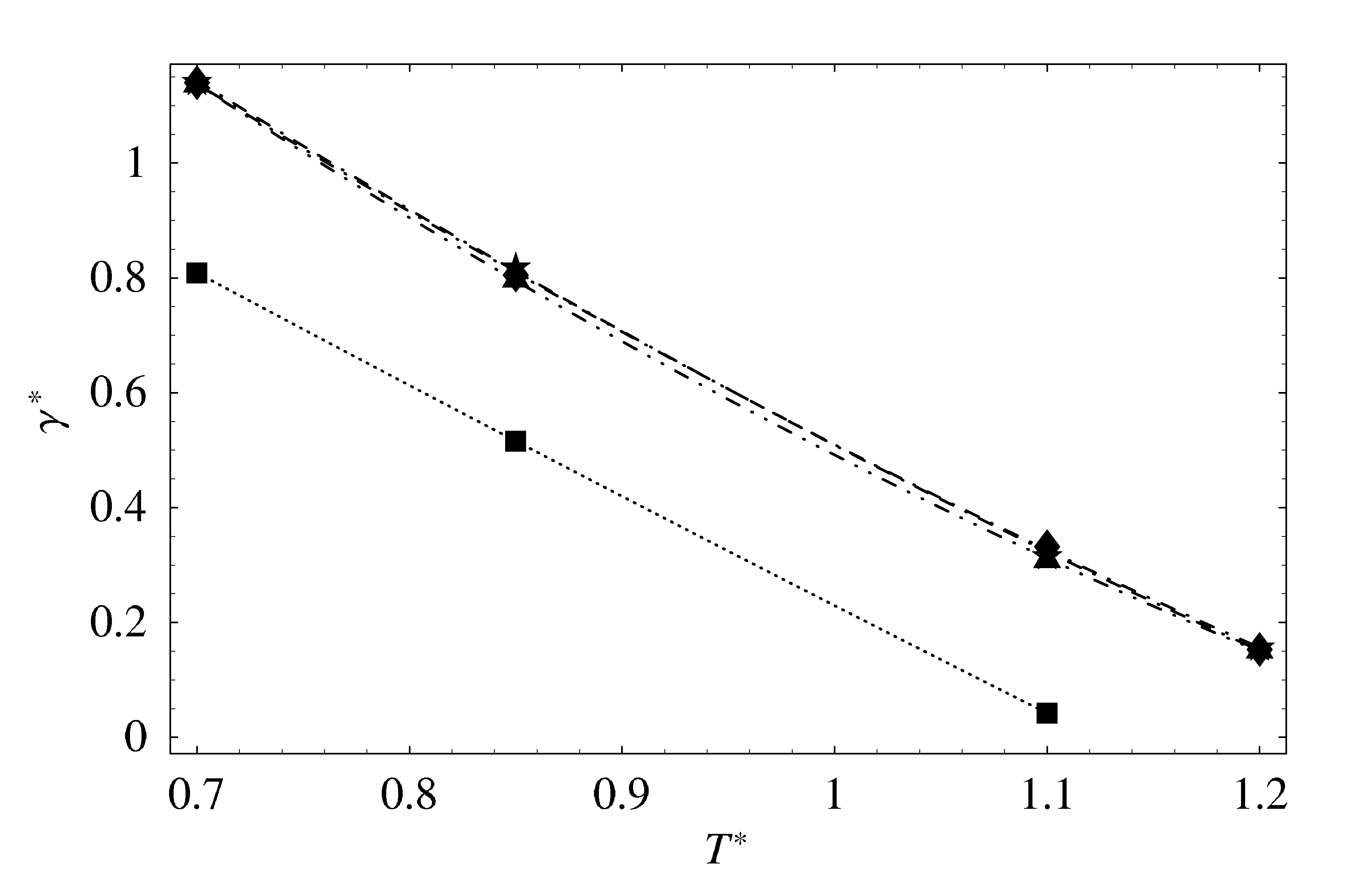}
\caption{Temperature dependence of the surface tension for a monomeric Lennard-Jones fluid at coexistence with cutoff $r_{c} = 2.5\sigma$ (square), $5.0\sigma$ (triangle), $7.5\sigma$ (diamond)  and Ewald summation crossover $r_{c}^{\mathbf{k}} = 5.0\sigma$ (star; overlaps with cutoff $r_{c} = 7.5\sigma$ data). 
Lines are added as a guide to the eye.
Reported surface tensions calculated by means of using a cutoff include tail corrections.\label{FigGamma}}
\end{center}
\end{figure}

\begin{figure}
\begin{center}
\includegraphics[width=3.2255in]{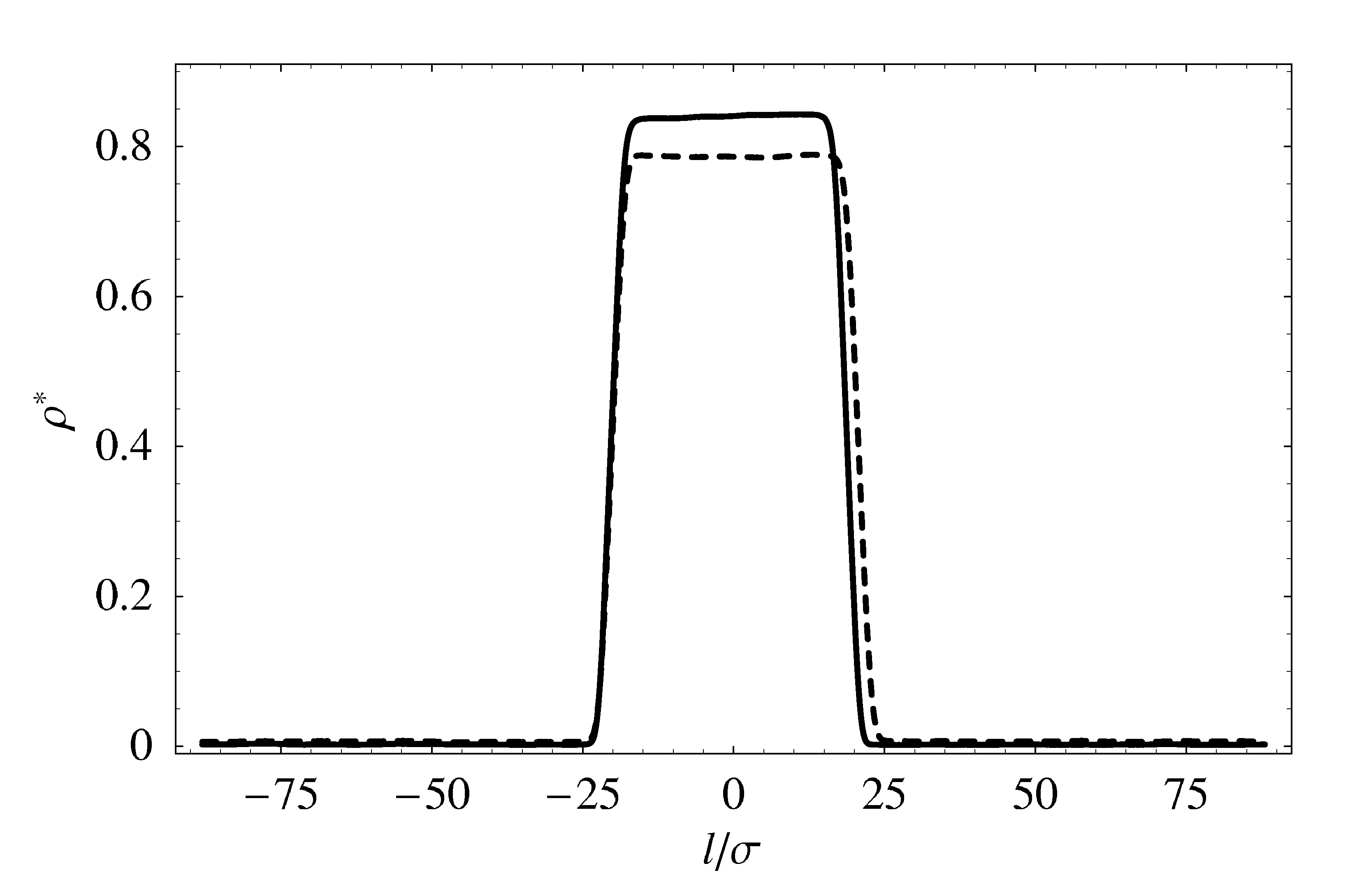}(a)
\includegraphics[width=3.2255in]{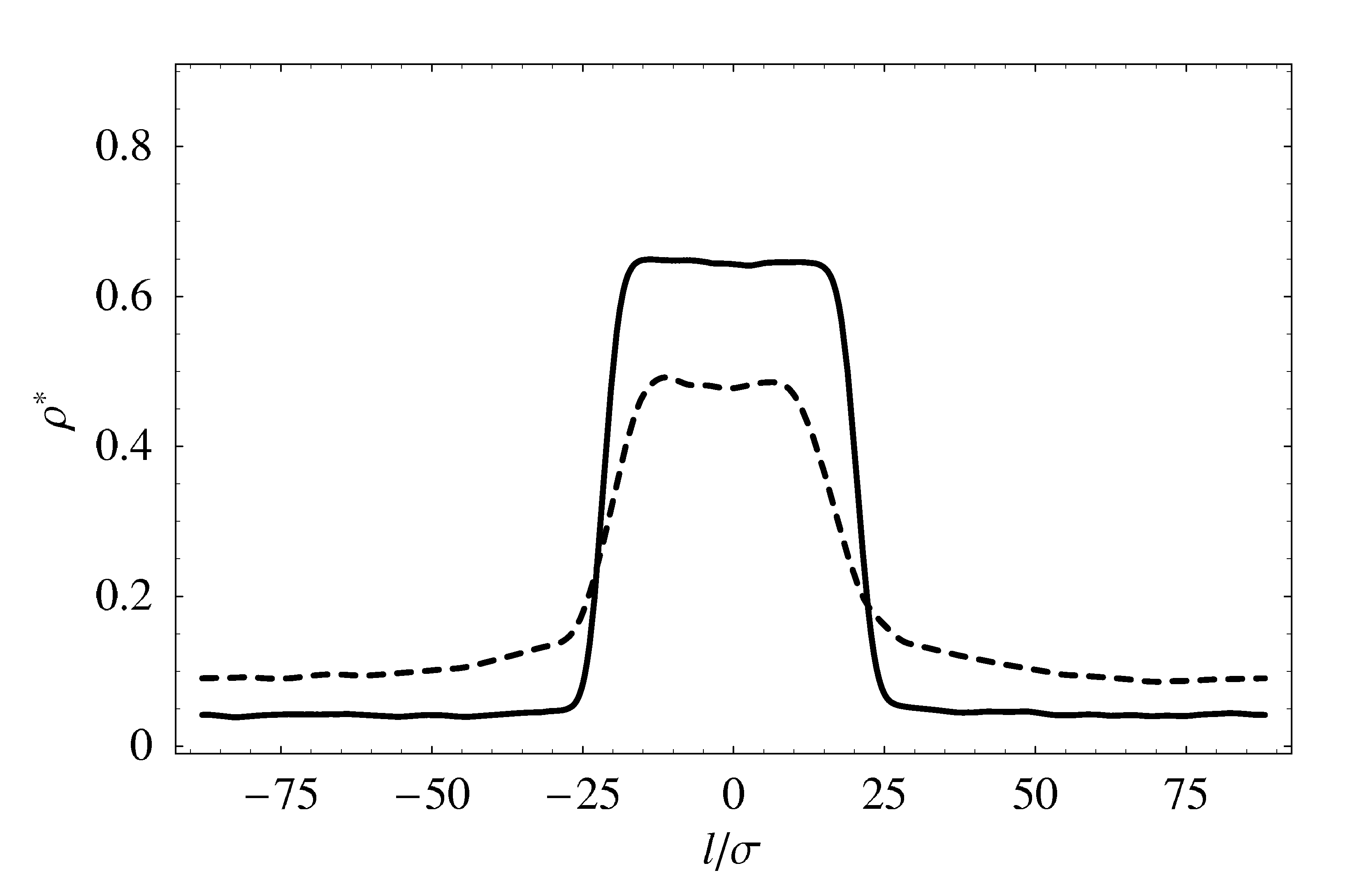}(b)
\caption{Density profiles of a monomeric Lennard-Jones fluid at coexistence with cutoff $r_{c} = 2.5\sigma$ (dashed) and Ewald summation crossover $r_{c}^{\mathbf{k}} = 4.0\sigma$ (solid) at a) $T^{*} = 0.70$ and b) $T^{*} = 1.10$.\label{FigDens}}
\end{center}
\end{figure}

The interfacial properties of the various LJ simulations are collected in Table \ref{TableLJ}.
The effects of using a short cutoff are immediately apparent from this table and Figure \ref{FigGamma}: at all temperatures studied, there is a significant difference between the calculated densities $\rho^{*}$, pressure $p^{*}$, and surface tension $\gamma^{*} = \gamma \sigma^{2}/\epsilon$ observed for a cutoff of $r_{\rm c} = 2.5\sigma$ and cutoffs of $r_{\rm c} = 5.0\sigma$ or $r_{\rm c} = 7.5\sigma$.
The liquid density of the $r_{\rm c} = 2.5\sigma$ simulation is statistically lower than the liquid densities at larger cutoffs; the vapor density is likewise statistically higher. 
The relative disagreement increases in size with increasing temperature, as can be seen in comparing the density plots of the $r_{\rm c} = 2.5\sigma$ and $r_{\rm c} = 7.5\sigma$ simulations for $T^{*} = 0.70$ and $T^{*} = 1.10$, as shown in Figure \ref{FigDens}.
The interfacial region is much broader for the smaller cutoff, indicating that the system is near its critical temperature $T^{*}_{\rm crit}$.
Literature confirms this observation: the critical temperature of Lennard-Jones fluids is approximately $T^{*}_{\rm crit} \approx 1.188$ \cite{Wilding95} for $r_{\rm c}=2.5\sigma$, which represents about an 11 percent decrease in the critical temperature when compared to $T^{*}_{\rm crit} \approx 1.316$ for $r_{\rm c}$ approaching infinity. \cite{Smit92}

Measured densities for Ewald dispersion simulations vary by not more than 2 percent for the liquid phase and 9 percent for the gas phase.
Furthermore, good agreement is observed between densities of simulations using cutoffs of $r_c = 7.5\sigma$ and corresponding densities obtained from Ewald dispersion simulations with $r_{\rm c}^{\mathbf k} = 5.0\sigma$.
Similar trends exist in the surface tension $\gamma^{*}$. 
Simulations using a cutoff yield surface tensions that at each temperature increase with the cutoff $r_{\rm c}$.
As the temperature increases, the observed values of $\gamma^{*}$ tend towards zero, which is another indicator that truncating the cutoff induces a lower critical temperature.

For simulations using Ewald summations, differences due to changes in crossover radii are much smaller.
In fact, for $T^{*} = 0.85$ and $T^{*} = 1.10$, there is essentially no difference between the values of $\gamma^{*}$ for crossover radii of $4.0\sigma$ and $5.0\sigma$; the results are identical within the uncertainty of the calculation.

Mecke {\em et al}. \cite{Mecke97} confirm our results when implicitly adding self-adjusting localized long-range corrections during simulation using on-the-fly density distributions instead of long-range correction through Ewald summations.
Their method requires a priori knowledge of the spatial distribution of the phases and breaks down for three dimensional phase distributions.
Application of Ewald dispersion summations circumvents this requirement and is applicable to any type of phase distribution.

\begingroup
\squeezetable
\begin{table*}
\newcolumntype{d}[0]{D{.}{.}{5}}
\caption{\label{TableLJ}
Interfacial properties of Lennard-Jones fluid$^{\rm a}$}
\begin{ruledtabular}
\begin{tabular}{@{\extracolsep{\fill}}lccccccccc}
Method &$N_{particles}$ & $r_{\rm c}/\sigma$ & $r_{\rm c}^{\mathbf{k}}/\sigma$ & $T^{*}$
& $\rho_{\rm liq}^{*}$ & $\rho_{\rm vap}^{*}$ 
& $\gamma^{*}_{\rm sim}$ & $\gamma^{*}_{\rm tail}$ & $\gamma^{*}$ 
 \\
\hline 
Cutoff &16000 &2.5 &--- &0.700(0) &0.7868(1) &0.0079(0) & 0.557(17) & 0.252 & 0.809 \\
&&5.0 &--- &0.700(0) & 0.8355(1) & 0.0026(0) & 1.033(17) & 0.104 & 1.137 \\
&&7.5 &--- &0.700(3) & 0.8401(1) & 0.0019(0) & 1.088(14) & 0.052 & 1.140\\
\cline{3-10}
&&2.5 &--- &0.850(0) & 0.7019(1) & 0.0274(0) & 0.331(05) & 0.183 & 0.516 \\
&&5.0 &--- &0.850(0) & 0.7675(1) & 0.0103(0) & 0.713(14) & 0.084 & 0.797 \\
&&7.5 &--- &0.850(1) & 0.7732(1) & 0.0100(0) & 0.762(08) & 0.043 & 0.805 \\
\cline{3-10}
&&2.5 &--- &1.100(0) & 0.4780(3) & 0.0847(2) & 0.026(11) & 0.016 & 0.042 \\
&&5.0 &--- &1.100(0) & 0.6383(1) & 0.0453(1) & 0.259(12) & 0.050 & 0.309 \\
&&7.5 &--- &1.100(0) & 0.6391(1) & 0.0514(1) & 0.304(08) & 0.028 & 0.332 \\
\cline{3-10}
&&2.5 &--- &1.200(0)$^{\rm b}$ & ---		     & 0.1330(1) & ---               & ---       & ---\\
&&5.0 &--- &1.200(0) & 0.5646(1) & 0.0781(1) & 0.132(12) & 0.019 & 0.151    \\
&&7.5 &--- &1.200(0) & 0.5648(1) & 0.0869(1) & 0.132(10) & 0.021 & 0.153    \\
\hline
Ewald &4000 &--- &3.0 &0.701(3) & 0.8340(1) & 0.0023(1)  & 1.060(61) & --- & 1.060 \\
& &--- &4.0 &0.700(3) & 0.8388(1) & 0.0023(1) & 1.066(42) &  --- & 1.066 \\ 
& &--- &5.0 &0.700(2) & 0.8400(1) & 0.0023(1) & 1.158(48) &  --- & 1.158 \\
\cline{3-10}
& &--- &3.0 &0.850(5) & 0.7663(1) & 0.0076(1) & 0.778(44) &  --- & 0.778 \\
& &--- &4.0 &0.851(5) & 0.7722(1) & 0.0078(1) & 0.817(32) &  --- & 0.817 \\
& &--- &5.0 &0.850(4) & 0.7742(1) & 0.0072(1) & 0.828(49) &  --- & 0.828 \\
\cline{3-10}
& &--- &3.0 &1.100(1) & 0.6353(2) & 0.0460(1) & 0.333(20) &  --- & 0.333 \\
& &--- &4.0 &1.100(1) & 0.6346(2) & 0.0413(1) & 0.373(38) &  --- & 0.373 \\
& &--- &5.0 &1.100(1) & 0.6452(2) & 0.0429(1) & 0.366(46) &  --- & 0.366 \\
\cline{3-10}
&16000 &--- &5.0 &0.700(0) & 0.8414(1) & 0.0020(0) & 1.141(11) &  --- &  1.141\\
& &--- &5.0 &0.850(0) & 0.7750(1) & 0.0098(0) & 0.818(15) &  --- &  0.818 \\
& &--- &5.0 &1.100(0) & 0.6383(1) & 0.0553(0) & 0.302(11) &  --- &  0.302 \\
& &--- &5.0 &1.200(0) & 0.5627(2) & 0.0942(0) & 0.156(14) &  --- &  0.156\\
\end{tabular}
\end{ruledtabular}
\begin{flushleft}
{$^{\rm a}$Digits between parentheses indicate the uncertainty in the last significant figure.} \\
{$^{\rm b}$Above the critical temperature.}
\end{flushleft}
\end{table*}
\endgroup

\subsection{SPC/E Water}

Our results for the water simulations are collected in Table \ref{TableW}. 
The most obvious difference between the results for Lennard-Jones fluids and those obtained for water is the difference in the relatively small importance of adding the long-range Lennard-Jones corrections to the potential for water.

We examined a number of different systems to analyze the effects of the $r^{-6}$ dispersion term, comparing systems with a cutoff to systems using Ewald summation for the dispersion sum.  
We first examined systems with uniform size and electrostatic crossover $r_{c}^{\mathbf k} = 10$ \AA; no clear trend was evident.
Keeping $L_{x}$ fixed while increasing both cutoffs uniformly likewise yielded no systematic behavior.
Increasing $r_{c}$ and $r_{c}^{\mathbf k}$ while keeping the ratio $L_{x}/r_{c}$ constant shows the expected increase in surface tension $\gamma_{\rm sim}$ with increasing cutoff. 
For the corresponding Ewald simulations, the surface tension $\gamma_{\rm sim}$ shows a much smaller variance with respect to changes in crossover $r_{c}^{\mathbf k}$ and system size.
Examining the overall surface tension, which incorporates the tail correction $\gamma_{\rm tail}$, we find that there is significantly greater variations in the simulations incorporating tail corrections for the $r^{-6}$ term than those using Ewald summations for the long-range interactions.

At 300 K, Ewald summation of the dispersion interaction yields a significantly improved estimate of the liquid-phase density $\rho_{\rm liq}$, particularly as the crossover distance $r_{\rm c}^{\mathbf{k}}$ increases.
By contrast, at 400 K, even for a cutoff of $r_{\rm c} = r_{c}^{\mathbf k} = 16$ \AA, the disagreement between the simulation and experimental densities is approximately 3 percent, which reflects the disagreement of the commonly used nonpolarizable water models with experimental results at temperatures well above 300 K.\cite{Taylor96,Jorgensen05}
However, differences between experimental and simulation results are significantly smaller when using Ewald summation than when employing a cutoff.


Comparing the total surface tensions $\gamma$ to the experimental results, however, we find that there is still a significant discrepancy between the two sets of measurements.
Although the inclusion of the long-range corrections yields an improved estimate compared to our previous work comparing the surface tensions of three- and four-point water models, \cite{Ismail06} the simulated surface tensions are still underestimated by approximately 10 to 15 percent. 
These results suggest that the SPC/E water model is incapable of accurately modeling interfacial properties, particularly at temperatures away from 300 K (the temperature at which the model was parametrized).

The relatively poor performance of the SPC/E water model may be an artifact of its original derivation, as the model is designed to perform well as bulk liquid, but is known to perform poorly when considering the vapor phase as a result of its dipole moment, which is significantly larger than its experimental value.
Such problems may be inherent to most non-polarizable models, which include most of the commonly used three- and four-point models;\cite{Ismail06}
polarizable models should perform better due to their adaptive nature.
However, our choice of the SPC/E water model was driven by its simplicity and the desire to illustrate the effects of a charged system on the performance of long-range dispersion summations.

\begingroup
\squeezetable
\begin{table*}
\caption{Interfacial properties of SPC/E water}
\begin{ruledtabular}
\begin{tabular}{@{\extracolsep{\fill}}cccccccccccc}
$T$ & Method  & $r_{\rm c}$ & $r_{\rm c}^{\bf k}$ & $N_{w}$ & $L_{x}$
& $\left<\rho_{\rm liq}\right>$ & $\left<\rho_{\rm vap}\right>$ & $\gamma_{\rm sim}$ 
& $\gamma_{\rm tail}$ & $\gamma$ \\
K$^{\rm a}$
 & & (\AA) & (\AA) &  & (\AA)  &(g/cm$^3$)$^{\rm b}$ & (g/cm$^{3}$) & 
 (mN/m)$^{\rm b}$
 & (mN/m)$^{\rm c}$ & (mN/m) \\
\hline
300 & Exptl.$^{\rm d}$  & --- & --- & & &  0.9965 & $3 \times 10^{-5}$ & --- & --- & 71.7 \\
& PPPM & 10 & 10 & 2200 & 36.5 & 0.9889 & $2 \times 10^{-5}$ & 55.2 & 6.4 & 61.6 \\
& & 12 & 10 & 2200 & 36.5 & 0.9924 & 0.0000 & 59.1 & 4.6 & 63.7\\
& & 12 & 12 & 2200 & 36.5 & 0.9909 & $3 \times 10^{-5}$ & 59.7 & 4.6 & 64.3 \\
& & 12 & 12 & 3200 & 43.8 & 0.9917 & $4 \times 10^{-5}$ & 57.1 & 4.6 & 61.7 \\
& & 16 & 10 & 2200 & 36.5 & 0.9954 & 0.0000 & 58.4 & 2.7 & 61.1 \\
& & 16 & 16 & 2200 & 36.5 & 0.9950 & $3 \times 10^{-5}$ & 56.1 & 2.7 & 58.8 \\
& & 16 & 16 & 5600 & 57.9 & 0.9940 & 0.0001 & 59.4 & 2.7 & 62.1 \\
& & 20 & 20 & 10000 & 77.4 & 0.9954 & $3 \times 10^{-5}$ & 59.6 & 1.7 & 61.3  \\
& Ewald & --- & 10 & 2200 & 36.5 & 0.9954 & 0.0006 & 61.8 & --- & 61.8 \\
& & --- & 12 & 2200 & 36.5 & 0.9956 & $4 \times 10^{-5}$ & 61.6 & --- & 61.6 \\
& & --- & 12 & 3200 & 36.5 & 0.9957 & 0.0000 & 61.1 & --- & 61.1 \\
& & 12$^{\rm e}$ & 12 & 3200 & 43.8 & 0.9912 & 0.0003 & 56.5 & 4.6 & 61.1 \\
& & --- & 16 & 2200 & 36.5 & 0.9966 & 0.0000 & 59.6 & --- & 59.6 \\
& & --- & 16 & 5600 & 57.9 & 0.9958 & 0.0001 & 60.5 & --- & 60.5 \\
\hline
400 & Exptl.$^{\rm d}$ & --- & --- & & &  0.9374 & 0.0014 & --- & --- & 53.6 \\
& PPPM & 10 & 10 & 2200 & 36.5 & 0.9076 & 0.0012 & 41.1 & 5.0 & 46.1\\
& & 12 & 12 & 3200 & 43.8 & 0.9106 & 0.0014 & 42.1 & 3.7 & 45.8 \\
& & 12 & 12 & 2200 & 36.5 & 0.9106 & 0.0015 & 42.5 & 3.7 & 46.2 \\
& & 16 & 16 & 2200 & 36.5 & 0.9138 & 0.0012 & 43.5 & 2.2 & 45.7 \\
& & 16 & 16 & 5600 & 57.9 & 0.9143 & 0.0012 & 40.9 & 2.2 & 43.1 \\
& Ewald & --- & 10 & 2200 & 36.5 & 0.9154 & 0.0014 & 43.8 & --- & 43.8 \\
& & --- & 12 & 2200 & 36.5 & 0.9154 & 0.0014 & 43.6 & --- & 43.6 \\
& & --- & 16 & 2200 & 36.5 & 0.9160 & 0.0009 & 43.0 & --- & 43.0 \\
& & --- & 16 & 5600 & 57.9 & 0.9165 & 0.0010 & 44.1 & --- & 44.1 
\end{tabular}
\end{ruledtabular}
\begin{flushleft}
{$^{\rm a}$Average temperatures agree with the thermostat temperature within 0.02 K.}\\
{$^{\rm b}$Uncertainties for densities are less than 0.002 g/cm$^{3}$; for surface tensions,  between 2.0 and 3.0 mN m$^{-1}$.}  \\
{$^{\rm c}$Estimated using tanh density profile.} \\
{$^{\rm d}$Data taken from Ref. \onlinecite{Lemmon05}.} \\
{$^{\rm e}$Long-range summation performed only for Coulombic interaction.}
\end{flushleft}
\label{TableW}
\end{table*}
\endgroup

\section{Algorithmic Performance\label{secP}}

While Ewald summations are the standard method for handling long-range forces in Monte Carlo simulations, for molecular dynamics simulations, Ewald summations are practical only for a relatively small numbers of particles.
Pollock and Glosli\cite{Pollock96} have shown  that for systems containing roughly 1000 particles or more, PPPM simulations outperform simulations that use explicit Ewald summation.
To reduce computational complexity, molecular dynamics simulations typically implement Fourier transform methods using meshes, such as the particle-particle particle-mesh (PPPM) method introduced by Hockney and Eastwood. \cite{Hockney88}
The PPPM method replaces the exact evaluation of the long-range interactions with an approximate treatment through the use of the fast Fourier transform over a discretized mesh.
This change reduces the complexity of computing the forces from $O(N^{2})$ to $O(N \log N)$ per time step.\cite{Deserno98}

Besides the number of particles, the computational performance of Ewald summations also depends on the number of $\mathbf{k}$-space vectors, which is derived from the chosen precision and crossover radius.
We compare performance of Ewald summations to cutoff simulations in two ways, using timing data reported in Table \ref{TableP}.
Comparisons between two simulations with equal number of particles and equal cutoff and crossover radius shows that Ewald summations are typically $2 \frac{1}{2}$ to 3 times slower.
However, the primary objective is to obtain results which approximate those that can be obtained without using a cutoff: 
we find simulations of 4000 particles with a crossover radius of $5.0\sigma$ have the same accuracy as simulations of 16000 particles with a cutoff radius of $r_{c} = 7.5\sigma$.
The timing comparison in this case shows that Ewald summations are about $1 \frac{1}{2}$ times faster.
Ewald summations have the added benefit of negating the need for post-processing of tail corrections.
This leads us to conclude that, regarding performance and accuracy, Ewald dispersion summations of Lennard-Jones simulations are preferable to using cutoffs combined with tail corrections. 

Although for smaller systems, Ewald summations are more efficient than mesh methods, as system size increases, which method is preferred can depend upon architecture- and problem-specific factors. 
This becomes even more evident when one considers problems also including Coulombic interactions, as is the case with water, as shown in Table \ref{table4}, which shows running times for a system of 6600 atoms on four processors.
The fastest running time for the Ewald simulations (357.0 ms) is significantly slower than the slowest running time for the PPPM simulation (308.6 ms), which uses a substantially larger cutoff radius (16 \AA\ versus 12 \AA).
It should be noted, however, that only temperatures well below the critical temperature were studied.
In general, we find that the PPPM simulation running time is dependent on the larger of the LJ cutoff and the crossover radius, while selection of an optimal Ewald crossover is complicated by the existence of an optimal crossover radius.
The slower execution of the Ewald algorithm, however, does not lead to a major improvement in accuracy: the difference between the surface tensions for the PPPM algorithm with tail correction and the Ewald algorithm for the various systems is roughly 1 percent.
Incorporation of the long-range dispersion interactions into a PPPM or other mesh-based formulation will be desirable for large systems.

\squeezetable
\begin{table*}
\caption{Running times for LJ simulations with cutoffs versus Ewald summations\label{TableP}}
\begin{center}
\begin{tabular}{@{\extracolsep{\fill}}ccccc}
\hline \hline
 &  & & \multicolumn{2}{c}{Time per timestep (ms)}
 \\
 \cline{4-5}
& & &  $N_{particles} = 4000$ 
 & $N_{particles} = 16000$ 
 \\ $T^{*}$ & Method & $r_{\rm c}$ ($\sigma$) & (4 procs) & (8 procs)
 \\
\hline
0.70 & Cutoff & 2.5 & ~~~5.2 & ~18.4\\
 & Cutoff & 5.0  & ~30.4 & ~40.6 \\
 & Cutoff & 7.5 & ~98.7 & 135.0\\
 & Ewald & 5.0  & ~95.1 & 124.6\\
\hline
0.85 & Cutoff & 2.5 & ~~~5.4 & ~~~6.7\\
& Cutoff & 5.0& ~33.1 & ~37.8\\
& Cutoff & 7.5& 115.1 & 129.0 \\
& Ewald & 5.0& ~87.7 & 121.9\\
\hline
1.10 & Cutoff & 2.5 & ~~~3.5 & ~~~6.4 \\
& Cutoff & 5.0& ~25.0 & ~35.9 \\
& Cutoff  & 7.5&  ~76.2 & 119.7 \\
& Ewald & 5.0 & ~67.3 & 103.5 \\
\hline \hline
\end{tabular}
\end{center}
\end{table*}%

\begin{table*}[htdp]
\caption{Running times (ms per timestep) for water simulations at 300 K using PPPM and Ewald summation for 6600 atoms on 4 processors\label{table4}}
\begin{center}\begin{tabular}{@{\extracolsep{\fill}}cccc}
\hline \hline
& \multicolumn{3}{c}{Crossover radius} \\
\cline{2-4}
Method &  $r_{c}^{\mathbf k} = 10$ \AA & $r_{c}^{\mathbf k} = 12$ \AA & 
$r_{c}^{\mathbf k} = 16$ \AA \\
\hline
PPPM, $r_{c} = 10$~\AA & ~93.3 & 143.8 & 289.4 \\
PPPM, $r_{c} = 12$~\AA & 140.9 & 144.8 & 296.4 \\
PPPM, $r_{c} = 16$~\AA & 277.7 & 292.4 & 308.6 \\
Ewald, $r_{c} = r_{c}^{\mathbf k}$ & 459.0 & 357.0 & 423.5 \\
\hline \hline
\end{tabular}
\end{center}
\end{table*}

\section{Conclusions}
\label{secC}

Simulated data shows the importance of long-range effects on pressure, density, and surface tension at all temperatures, but especially close to the critical point.
For simulations of a homogeneous bulk system, cutting off the dispersion term at $r_c$ is sufficient provided that the pair correlation function $g(r)$ for $r>r_c$ is close to 1.0. In this case, the forces for distances $r>r_c$ cancel, and the contributions to the pressure and energy for $r>r_c$ can be easily accounted. 
However, for two-phase systems such as those studided here, cutting off the dispersion contribution to the pair potential have strong effects on the properties of the system. 
Part of the missing contribution to the surface tension can be included by using tail corrections, but since truncating the potential at $r_c$ can have a strong effect on the liquid and vapor densities, particularly near the critical point, inclusion of the tail correction for the surface tension only accounts for part of the missing contribution.
This supports the notion that description of long-range effects by
means of Ewald summation is preferential close to the critical point.
The long-range effects become less pronounced when further removed from the
critical point.
Comparison between the Lennard-Jones fluid and water results show
that the importance of long-range dispersion effects become less pronounced
when stronger contributions, such as Coulombic terms, are included.
In this case, the latter is the dominant contribution, while the former plays
a secondary role.

\section*{Acknowledgments}

Sandia is a multiprogram laboratory operated by Sandia Corporation, a Lockheed Martin
Company, for the United States Department of Energy under Contract No. DE-AC04-94AL85000.

\bibliographystyle{aip}

\end{document}